# Addressable Graphene Encapsulation of Wet Specimens on a Chip for Combinatorial Optical, Electron, Infrared and X-ray based Spectromicroscopy Studies

*Christopher Arble[1†], Hongxuan Guo[2†], Alessia Matruglio[3†], Alessandra Gianoncelli[4], Lisa Vaccari[4], Giovanni Birarda[4] and Andrei Kolmakov[1*]*

[1]Physical Measurement Laboratory, National Institute of Standards and Technology, Gaithersburg, MD 20899, USA

[2]SEU-FEI Nano-Pico Center, Key Laboratory of MEMS of Ministry of Education, Southeast University, Nanjing 210096, P. R. China

[3]CERIC-ERIC (Central European Research Infrastructure Consortium), S.S. 14 Km 163,4 in Area Science Park, 34149, Basovizza, Trieste, Italy

[4]Elettra Sincrotrone Trieste S.C.p.A., S.S. 14 Km 163,4 in Area Science Park, 34149, Basovizza, Trieste, Italy

[†] These authors contributed equally

[*] Correspondence should be sent to andrei.kolmakov@nist.gov

ABSTRACT:

Label-free spectromicroscopy methods offer the capability to examine complex cellular phenomena. Electron and X-ray based spectromicroscopy methods, though powerful, have been hard to implement with hydrated objects due to the vacuum incompatibility of the samples and due to the parasitic signals from (or drastic attenuation by) the liquid matrix surrounding the biological object of interest. Similarly, for many techniques that operate at ambient pressure, such as Fourier Transform Infrared spectromicroscopy (FTIRM), the aqueous environment imposes severe limitations due to the strong absorption by liquid water in the infrared regime. Here we propose a microfabricated multi-compartmental and reusable hydrated sample platform suitable for use with several analytical techniques, which employs the conformal encapsulation of biological specimens by atomically thin graphene. Such an electron, X-ray, and infrared transparent, molecularly impermeable as well as mechanically robust enclosure preserves the hydrated environment around the object for a sufficient time to allow *in-situ* examination of hydrated bio-objects with techniques operating both in ambient or high vacuum conditions. An additional hydration source, made by hydrogel pads patterned near/around the specimen and co-encapsulated, has been added to further extend the hydration lifetime. Scanning electron and optical fluorescence microscopies as well as synchrotron radiation based FTIR and X-ray fluorescence microscopies have been used to test the applicability of the platform and for its validation with yeast, A549 human carcinoma lung cells and micropatterned gels as biological object phantoms.

## 1. Introduction

The spectromicroscopy of hydrated cellular specimens, bacteria, viruses, and microorganisms under physiological conditions is an attractive tool for better understanding realistic biological processes. However, preserving the wet environment near the cells during an experiment presents a challenge for many analytical techniques such as Scanning /Transmission Electron Microscopies (SEM, STEM, TEM), Scanning Transmission /Photoelectron X-ray microscopies (STXM, SPEM), Photoemission Electron Microscopy (PEEM), and in general for those techniques that require vacuum for signal acquisition. These challenges also exist for techniques that operate under ambient pressure, such as Fourier Transform Infrared spectromicroscopy (FTIRM), and those which require minimal influence of water matrix on sample spectra. The very early examples of the electron-based microscopy and spectroscopy application in aqueous conditions have been shown already early in the last century.[1-3] Since then, there have been many studies of biological constituents using electron spectroscopy and microscopy techniques employing either a differentially pumped sample enclosures or ultra-thin SiN[4-7] and other electron transparent membranes[8-11] that separate wet sample from vacuum (see also Ref.[12] and corresponding chapters therein). However, the significant attenuation of primary, secondary or photoelectrons by water molecules imply severe requirements on the water layer and entire sample thicknesses along the probing direction. As an example, a liquid sample only a few nanometers thick can be probed if low energy (10 eV to 1000 eV) (photo-) electrons are used for analysis.[13] Comparable considerations hold true for soft X-ray spectromicroscopy.[14] For instance, the water window is of particular interest for STXM microscopy of unstained biological samples since it represents an electromagnetic spectral region (282 eV to 533 eV) where water is highly transparent while carbon possesses a high absorption cross-section. Yet, in a standard liquid cell design, where the sample is confined between two SiN membranes, only a very thin water layer is allowable along the optical path to reduce attenuation of photons for transmission X-ray imaging or for X-ray fluorescence elemental analysis. Similarly, sandwiching the sample between two $CaF_2$ windows is routinely used in biocompatible fluidic devices for non-ionizing FTIRM of live cells.[15] Water, however, is a strong IR-absorber itself, and a straightforward interpretation of cellular IR spectral features is often impeded due to overwhelming signatures of the surrounding aqueous media.[16] In part those challenges can be mitigated using Attenuated total reflection FTIR (ATR-FTIR), which, however, also limits the penetration depth of IR light in the sample to hundreds of nanometers[17], shorter than the thickness of an hydrated cell.

The plausible solution is a conformal encapsulation of biological objects with a thin flexible membrane. This would provide a minimal amount of aqueous media along the optical path, thus reducing the absorption and scattering by the aqueous matrix while still providing cells with a hydrated environment for an extended period of time sufficient for spectroscopic analysis.[9] Along this line, the progress in fabrication and handling of two dimensional (2D) materials has enabled the fabrication of ultrathin electron and photon transparent conforming biocompatible enclosures.[18] Graphene membranes provide several unique advantages as a window material, namely they are molecularly impermeable, atomically thin and flexible while being the strongest possible encapsulating material. In addition, graphene membranes are electrically and thermally conductive thus eliminating sample charging and reducing local heating. The graphene enclosure isolates and maintains a hydrated sample environment not only in the ambient but also under high- and ultra-high vacuum (HV/UHV) conditions for extended periods of time enabling access to hydrated samples with a wide range of useful electron spectroscopy and microscopy instrumentation.[19-23]

With respect to the analysis of biological samples, the encapsulation by graphene has been already widely recognized. Mammalian cells were imaged via electron microscopies, scooping up a single graphene layer directly onto the specimens, to trap some water at the graphene-support interface and maintain that the samples were hydrated and separated from the ambient environment.[24] Other work has shown the three dimensional dynamics of double strand DNA conjugated with Au nanocrystals imaged using TEM by encapsulating the specimen between two single graphene layers sealed by van der Waals forces.[25] Infrared near-field nano-spectroscopy of the wet tobacco mosaic virus was demonstrated using a single layer of graphene as an impermeable cover, allowing for imaging of individual viruses.[26] Moreover, TEM and fluorescence light microscopy of hydrated Madin−Darby canine kidney epithelial cells was performed via transferring multilayer free standing graphene on cells grown on a graphene-coated TEM grid.[11] Recently, live bacteria imaging was obtained using SEM on *Escherichia coli, Lactococcus lactis,* and *Bacillus subtilis* deposited on a silicon substrate and encapsulated using a single layer graphene veil. The graphene layer maintained a hydrated environment, allowing the study of metabolic phenomena and intracellular interactions, such as fission and apoptosis.[27]

In previous work, we have already shown how graphene can be used as conformable water confining membrane for FTIR imaging of hydrated biological cells. In particular it was shown that the layer of water confined by graphene is sufficiently thin to not interfere with the cellular spectral features, allowing a clear spectral interpretation.[28] Nevertheless, we found that the commonly adopted graphene transfer method, based on the so called "scooping up" approach,[29] is not particularly applicable for encapsulation of biological objects since it implies immersion of the sample into harsh solutions, such as acetone, when removing the PMMA supporting layer. On the other hand, we found that the scooping of unsupported graphene out of water is not reliable as graphene breaks into flakes of uncontrollable size. Therefore, there exists a need for a simple, reliable, and biocompatible graphene encapsulation process.

Increasing the liquid retention time inside the graphene conformal enclosure is another experimental challenge. In this respect, accompanying the graphene encapsulation of the cells with a biocompatible on-board source of liquid would be desirable. Hydrogel scaffolds are highly porous and biocompatible materials, that have been routinely used for drug delivery, tissue engineering, biosensing, wearable electronics and soft robotics.[30] Immobilization of small living organisms such as the nematode *Caenorhabditis elegans* via photo-crosslinking of the surrounding gel solution has been already reported.[31] The addressable cell immobilization with a patternable gel media could allow extended cellular lifetime, via supplying nutrients and feedstocks, during measurement and locking of the mobile samples.[32] Immobilization of cells with crosslinking hydrogels was shown to be a suitable method that provides the key conditions of hydration, nutrition, material exchange, and pH, while still providing minimal spectral interference and damage from low dose ionizing radiation.[32] As a matter of principle, hydrogels can further improve the performances of graphene encapsulation and contribute the extension of the sample life and the reliability of the measurements.

Herein we report on the fabrication and validation testing of a graphene encapsulation liquid cell (GrELC) designed to be suitable for *in-situ* studies of biological samples with an array of analytical methods such as optical/fluorescence/infrared as well as scanning electron and X-ray microscopies. This is a multicompartmental (3x3 array) platform where the samples can be placed and immobilized in an addressable way on 50 nm thin SiN windows and become encapsulated by the graphene membrane once a special lid is put in place. The design relies on the construction of a flexible polymer lid with multiple graphene covered viewports and therefore decouples the graphene scooping and sample encapsulation procedures. The test samples used in the present work are mammalian and yeast cells and hydrogel microstructures. More specifically, Polyethylene Glycol Diacrylate (PEGDA), an UV/electron beam patternable biocompatible hydrogel, was exploited in this study either as hydrated biological object phantom or as a co-encapsulant additional hydration source. Specifically, the integration of the hydrogel patterning in the design of the GrELCs further increases the lifetime of the device in a hydrated state, opening exciting possibilities for a new liquid retention concept, based on object encapsulation together with medium supply. The sample platform proposed herein facilitates the study of biological samples, under physiological conditions with techniques that require minimal influence of water matrix on sample signal and/or HV conditions.

## 2. Materials and methods

### 2.1 Graphene Encapsulation Liquid Cell Design

The GrELCs was designed and fabricated using the facilities at the Center for Nanoscale Science and Technology at NIST (Gaithersburg, USA). It consists of two parts: (i) Si/SiN/SiO$_2$ sample support chip with a 3x3 array lithographically defined 50 nm thick and 100 µm x 100 µm wide SiN windows placed over 100 µm in diameter orifices in 1 µm thick SiO$_2$ layer; and (ii) flexible 25 µm or 6 µm thick lid made of SU-8 developed photoresist that has a corresponding 3x3 array of 100 µm holes (viewports) covered with 2 to 4 layers of CVD grown graphene using standard PMMA based transfer method[29] . The concept and the design of the GrELC are shown in Figure 1. Figures S1 and S2 in Supporting Information file provide further information on the fabrication of the two parts of the chip: the support and the lid.

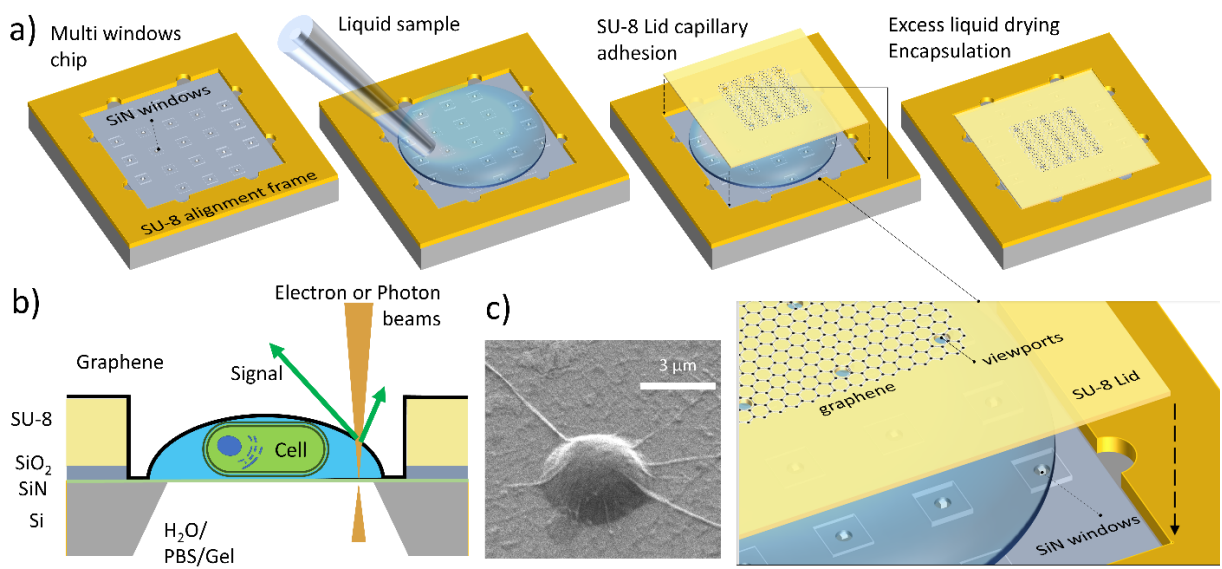

Figure 1 The concept and design of two parts graphene encapsulation liquid cell, GrELC. a) The sample filling and assembly of the GrELC; b) the general concept of GrELC for a wide range of spectromicroscopies; c) Graphene encapsulated yeast cell; SEM 2 keV. The right bottom panel represents enlarged top panel with graphene covered SU-8 viewports of the lid being co-aligned with SiN windows by SU-8 frame.

The samples of interest (*e.g.* procaryote and eukaryote cells) are drop casted on to the multi-window chip and selectively (see description below) or randomly adhered to the SiN windows. The SU-8 lid with graphene covered windows is applied to the liquid filled cavity and adheres evenly by capillary forces. The SiN windows and graphene covered viewports are precisely aligned by the SU-8 frame on the main chip and excess liquid can be controllably removed with a filter paper. The remaining interlayer liquid slowly evaporates and leaves behind pre-designed areas of solution trapped by the graphene membrane. This process can be observed under an optical microscope and occurs within minutes. This two-piece platform design eliminates the risk of exposure of sensitive samples to reactive solutions common in standard encapsulation procedure via scooping process. The multiple viewports can be used for comparative-combinatorial imaging and analysis. The parts are, in principle, reusable by removing graphene and samples with solvents and plasma ashing.

**2.2 Addressable immobilization of live cells and microfabrication of model objects using hydrogels**

The addressable immobilization of live cells on SiN windows has been performed via two different methods. Mammalian cells can be selectively grown prior to the experiment using protocols described in section 2.4 below. Alternatively, a hydrogel patterning procedure was developed to immobilize live yeast cells with a biologically compatible material in a controllable way exploiting a patterning method with micron scale precision.[32] To achieve this, a mixture of the biocompatible 20 % w/v PEGDA solution in Phosphate Buffer Solution (PBS) and yeast cells was applied to the front side of the multi-window chip and crosslinked either by 365 nm UV light or an electron beam from the back side through the SiN window to solidify the hydrogel solution around the object (Figure 2a). The thickness (and porosity) of the crosslinked hydrogel depends on the radiation attenuation by the media and dose of the irradiation and can be tuned from ≈ 50 nm to ≈ 50 µm. As UV light is diffusive, the immobilization of the samples takes place uniformly across the entirety of the SiN window. On the contrary, the use of an electron beam of SEM allows for the programmable patterning of high-fidelity gel structures of different shapes and thickness. The encapsulation tests were carried out with *saccharomyces cerevisiae* yeasts cells as a model bio-object (see the section "Yeast cells preparation" in SI). Alternatively, hydrogel structures of micron size can mimic hydrated biological objects. The latter, in conjunction with optical fluorescence microscopy were used to evaluate the water retention time inside the encapsulated objects. In either application, the chip with gelated structures was rinsed with DI water or PBS and then sealed into the GrELC (Figure 2 b-d).  The exemplary SEM and STEM images of the graphene encapsulated yeast cells in contact with hydrogel structure are shown in Figure 2 e-g. Additional images and information are provided in SI-file and in the Figure S3.

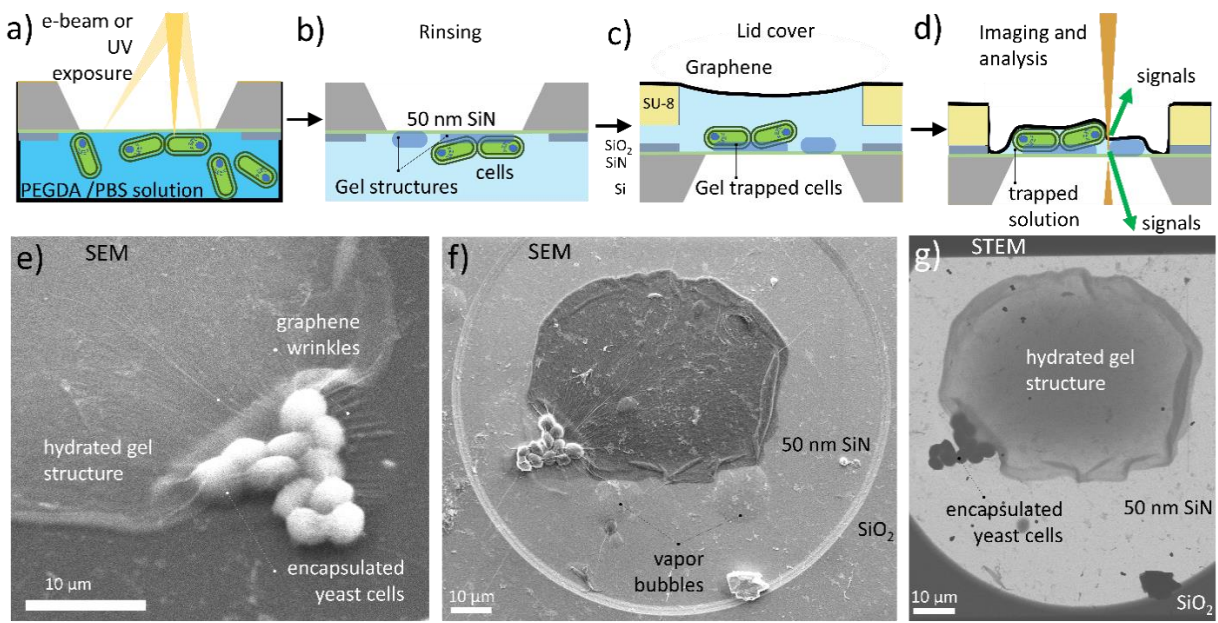

**Figure 2** Addressable immobilization of live cells with hydrogel followed by graphene encapsulation. a) Irradiation of composite PBS solution with cells and PEGDA molecules through SiN membrane leads to formation of the gel layers and structures which can trap the live cells. b-d) After rinsing and covering the multiwindow chip with graphene lid. The gel-immobilized graphene encapsulated cells have an additional hydration source compared to pristine encapsulated cells. e) SEM image of an exemplary graphene encapsulated hydrated gel structure with trapped test cells; f) Low magnification SEM image of the $SiO_2$ cavity with SiN membrane containing graphene encapsulated yeast cells and gel layer. The presence of water vapor bubbles proves the hydrated state of the gel; g) The same object imaged in transmission mode 30 keV STEM. The darkening of the gel toward the center presumably follows the gradient of water content.

## 2.3 Scanning Electron Microscopy

SEM imaging and EDS measurements were conducted in a field emission scanning electron microscope. An Everhart–Thornley (E-T) detector was used for SEM imaging which detects both secondary and backscattered electrons. The base pressure of the instrument during acquisition was $1 \times 10^{-4}$ Pa. SEM imaging of graphene was usually taken at electron beam energy 2 KeV to enhance hydrated vs dry sample contrast and reduce water radiolysis. For SEM imaging in transmission mode the electron energy was increased to 30 keV. The gel feature writing and immobilization of the cells with crosslinking PEGDA gel was conducted by irradiation the solution through SiN membrane with 5 KeV, 10 KeV and 15 KeV electron beams for several seconds such that the total accumulated dose did not exceed 10 e$^-$/nm$^2$.[32] For EDS mapping, a 5 keV, 160 pA electron beam was used as a compromise between S/N ratio and radiolitic damage of the sample. SEM and EDS measurements were collected on wet GrELCs as a function of time to determine dehydration rate under high vacuum conditions (see results below). The beam was blanked between measurements.

**2.4 Fourier Transform Infrared Microscopy - FTIRM**

Human lung carcinoma A549 cells were grown directly on the SiN membranes prior to the experiments. Before plating cells, membranes were sterilized under UV light for 3 hours and then washed with EtOH 70 % v/v. Cells were grow in Dulbecco's modified Eagle's medium (DMEM) supplemented with 10 % FBS at 37 C in 5% $CO_2$ atmosphere, refreshing the media every 24 hours. After four days, cells were fixed in paraformaldehyde (PFA) 3.7 % in PBS. After cells fixation, the lid with graphene is placed onto the chip with the cells. The graphene conforms over the specimen confining a thin liquid layer and the excess of solution is expelled laterally.

FTIRM measurements were carried out at SISSI beamline - Chemical and Life Sciences branch at Elettra Sincrotrone Trieste, Italy[33], a few minutes after graphene encapsulation process. FTIR measurements were performed by using an in-vacuum interferometer coupled with a nitrogen-flushed Vis-IR microscope, equipped with a liquid nitrogen-cooled Mercury-Cadmium-Telluride (MCT) detector. Single point IR spectra of individual A549 cells were collected at 20 µm x 20 µm$^2$ spatial resolution from 4000 cm$^{-1}$ to 800 cm$^{-1}$ in double side, forward/backward acquisition mode with a scanner velocity of 120 kHz. 512 scans were averaged with and the spectral resolution was 4 cm$^{-1}$. The acquired raw data were corrected for water vapor and $CO_2$ absorption during data analysis process.

**2.5 X-ray Fluorescence Spectroscopy**

Elemental distribution maps of A549 human lung cells were obtained by performing low-energy X-ray fluorescence (XRF) measurements at the soft X-ray microscopy TwinMic beamline at Elettra.[34] The beamline was operated in scanning mode at a photon energy of 1.03 keV. The incoming monochromatized X-ray beam was filtered through an aperture with a diameter of 75 µm, acting as secondary source. Exploiting a Fresnel focusing zone plate (FZP) with a diameter of 600 µm and an outermost zone width of 50 nm, the X-ray beam was focused at the sample position, 24.2 mm downstream the FZP, obtaining a focal spot size of 0.88 µm in diameter. A scintillator-based CCD imaging system[35] was located further downstream to collect the transmitted signal, providing morphological information via absorption and differential phase contrast. Eight radially-distributed silicon drift detectors (SDD) were positioned 28 mm upstream and at an angle of 20° with respect to the sample plane and were used to collect X-ray fluorescence photons emitted by the specimen.[36] For each region of interest (ROI), the sample was scanned across the beam with a step size of 800 nm and a dwell time of 5 sec/pixel for XRF measurements and 20 ms for the transmission imaging. The recorded XRF spectra were fitted using energy dispersive X-ray fluorescence data processing software,[37] producing elemental distribution maps relative to the characteristic fluorescence emission K lines of oxygen.

### 3. Results and Discussion

Below we describe few application examples of GrELC's aimed to highlight their versatility with respect to three spectromicroscopy techniques, namely FTIRM, Low Energy (LE)-XRF and SEM-EDS. Depending on the method used, we considered two different sample classes: i.e. graphene encapsulated pristine cells with small amount of encapsulated water, and graphene-encapsulated gel-immobilized cells. The latter technology was introduced to increase liquid retention time under both environmental and vacuum conditions. Since a unified definition of the liquid retention time across the applied techniques is not feasible, due to peculiarity of each analytical technique and their diverse sensitivity, an explanation of technique- specific retention time will be provided along the discussion.

To demonstrate the performance of the GrELC design, A549 human carcinoma alveolar cells were analyzed by FTIR microscopy and LE STXM-XRF, two forefront techniques for life sciences studies, often used in conjunction thanks to the complementarity of the information they provid.[14, 16, 38, 39] Since FTIRM probes the vibrational modes of molecular moieties and LE STXM-XRF is especially sensitive to light elements, such as C, O, and N, any low Z encapsulating matrix potentially interferes with the chemical information from the sample. For this reason, we used pristine cells instead of PEGDA immobilized cells. Furthermore, since the design of the present GrELC is static (not fluidic), aiming primarily to test the long-term hydration capacity of the GrELC, we have opted to work with hydrated formalin fixed-cells, to avoid misinterpretation of the spectral features eventually deriving from cell autolysis, i.e. enzymatic self-digestion due to osmotic stress and nutrients deprivation.[40] With respect to other cell fixation methods, such as ethanol dehydration, formalin fixation is well-established to minimally perturb the spectral profile of mammalian cells.[41] It is therefore not surprising that the FTIR absorption spectrum in the mid-infrared region of hydrated formalin-fixed A549 cells in the GrELC, shown in Figure 3 (A549 GrELC, green spectrum) looks very similar to the spectrum of the same formalin-fixed dried cell grown on SiN membrane (Figure 3, A459 dried SiN blue spectrum). Conversely, one can see apparent differences with respect to the infrared spectrum of a whole A549 cell collected in conventional $CaF_2$ liquid cells (Figure 3, A549 $CaF_2$ liquid cell, black spectrum). The latter one is dominated by liquid water spectral modes, and especially by the -OH stretching modes in the 3800-3000 $cm^{-1}$ spectral region, as well as by the H-O-H bending mode centered at about 1640 $cm^{-1}$, as highlighted by the shadowed blue region in Figure 3a.[42] This strong overlap impedes the spectral interpretation and often severely limits it.[43] Conversely, A549 GrELC cell spectrum, being not affected by the constraints imposed by thick water overlayers common to conventional hydrated cell spectra, can be easily interpreted on the base of the prior works. This observation brings the attention to another relevant point: liquid water is a very strong IR absorber[44] and the fact that its features are not immediately detectable highlights that the encapsulation is made by a very thin layer. Considering A549 GrELC cell spectrum, methyl and methylene asymmetric and symmetric stretching modes associated to aliphatic chains, mostly of cellular lipids, are clearly visible in the 3000-2800 $cm^{-1}$ spectral region. These components are barely accessible in A459 $CaF_2$ liquid cell spectrum, since they are overwhelmed by tail of the –OH stretching mode of liquid water.[42] Indeed, in this spectral region Amide A band at about 3300 $cm^{-1}$, localized on the NH group of the peptide linkage of cellular proteins, can be also clearly distinguished in the A549 GrELC cell spectrum, while it is completely not accessible in A549 $CaF_2$ liquid cell spectrum. In addition, the Amide I band centered at ≈1667 $cm^{-1}$, mostly associated to the carbonyl stretching of the peptide

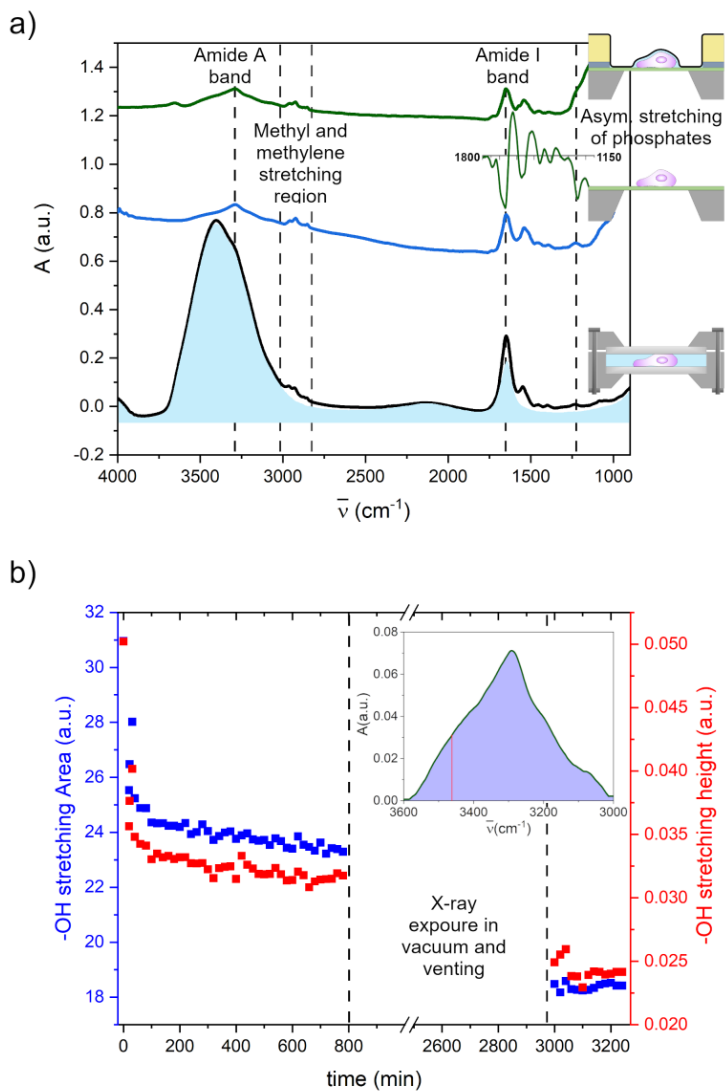

**Figure 3** a) Black line: FTIR spectra of formalin fixed hydrated A549 cell in a conventional CaF$_2$ liquid cell (as sketched on the right of the spectrum). The water contribution to the cell spectrum is highlighted as shadowed blue region; Blue line: FTIR spectrum of formalin fixed dried A549 cell on SiN substrate. The sampling geometry is sketched on the right of the spectrum; Green line: FTIR spectra of formalin fixed hydrated A549 cell in GrELC (sampling geometry sketched on the right of the spectrum). Inset: second derivative of GrELC cell spectrum in the 1800-1100 cm$^{-1}$ spectral region; b) Time evolution of the water content in the GrELC cell estimated as area integral of the -OH stretching modes between 3600-3000 cm$^{-1}$ (blue squares), as well as considering the height of the band at 3460 cm$^{-1}$ (red squares).

backbone of cellular proteins, is also easily detectable in the A549 GrELC spectrum. This band, the relevance of which for the assessment of the protein secondary structure is well documented, is also severely affected by the water bending mode in A549 CaF$_2$ liquid cell spectrum, and the need of complex, tedious, and unavoidably inaccurate post-processing water subtraction procedures for its disclosure makes questionable any fine spectral attribution. Finally, the asymmetric stretching of phosphates, centered at about 1245 cm$^{-1}$, and associated to nucleic acid backbone and phospholipids heads,[45] is barely visible as a shoulder of the strong absorption associated to SiN stretching modes, while it can be seen in a second derivative spectrum (see Figure 3a inset). Nevertheless, the full access to this spectral band, as well as to the symmetric phosphate stretching mode, usually centered at about 1085 cm$^{-1}$, can be improved by using SiC membrane supports, as already proven by the authors,[46] without changing the design and fabrication steps of GrELC here proposed for SiN.

In order to evaluate the sealing efficiency of the GrELC, fifteen A549 cells have been cyclically measured for about 14 hours, and the water retention time was evaluated on the base of the water spectral content, estimated both as area integral of the 3600 cm$^{-1}$ to 3000 cm$^{-1}$ spectral region, as well as the height of the band centered at about 3460 cm$^{-1}$, associated to one of the components of the -OH stretching band of water.[47] The trend for the exemplary A549 cell is shown in Figure 3b, and the inset of the same figure highlights both the selected area and peak height. During the first measurement hour there is a clear water evaporation trend, which can be attributed to the evaporation of the excess of water from the graphene-SiN interface to the ambient; then the integral values remain nearly stable for the rest of the measurements. After about 14 measurement hours at SISSI-Bio beamline, the very same GrELC has been transferred to the vacuum chamber of TwinMic microscope (pressure ≤ 1 × 10$^{-5}$ mbar ≈ 10$^{-3}$ Pa) for XRF mapping of the A549 cells.

Figure 4a shows a representative STXM absorption image of A549 cells grown on SiN membranes: as expected, they grew as an adherent monolayer of pebble-like shaped cells.[48] Figures 4 b-h depicts the oxygen distribution in a representative GrELC, as revealed by LE-XRF, at time t = 0 (panel b), immediately after having reached the operational pressure of TwinMic microscope, and at 2.7 hours, 5.2 hours, 7.7 hours, 10.4 hours, and 17 hours in vacuum for the panels b) to f) respectively. A549 cells are clearly distinguishable from the LE-XRF oxygen maps (black arrows in panel b): the higher oxygen content being at nuclear level, decreasing in the perinuclear and cytoplasmic sample compartments. The oxygen content in the cell-surrounding areas is lower: it slightly decreases in the first 5 hours of measurements in the intercellular regions (white arrows in panels b to d), while remaining almost constant for the following several hours in vacuum (panels e to g). The map h) was instead acquired after venting the vacuum chamber for GrELC visual inspection. Indeed, we verified that the venting procedure induced some cracking on the graphene and therefore it is plausible that the map f), collected after having reached the TwinMic standard measurement conditions ($\approx 10^{-3}$ Pa), was taken on a fully dried sample. It reveals a clearly detectable decrement of the oxygen level both a cellular and intercellular level. The observed time dependent oxygen trend can be interpreted assuming that water encapsulated by graphene, the content of which can be estimated considering the oxygen signal decrement, tightly follows the cellular profile and surrounding cellular areas, as sketched in Figure 2d. The chamber vacuum induces possibly more water redistribution by interfacial diffusion (see later for more details) than water evaporation within the time the sample has been kept in vacuum before venting, while clear evidences of water evaporation could be detected only after the vacuum beak and consequent graphene layer rupture. The described trend is fully consistent with FTIR analysis. In Figure 3b, the clear intensity spike of the selected water diagnostic modes before and after X-ray exposure can be easily appreciated and fully in accordance with LE-XRF data.

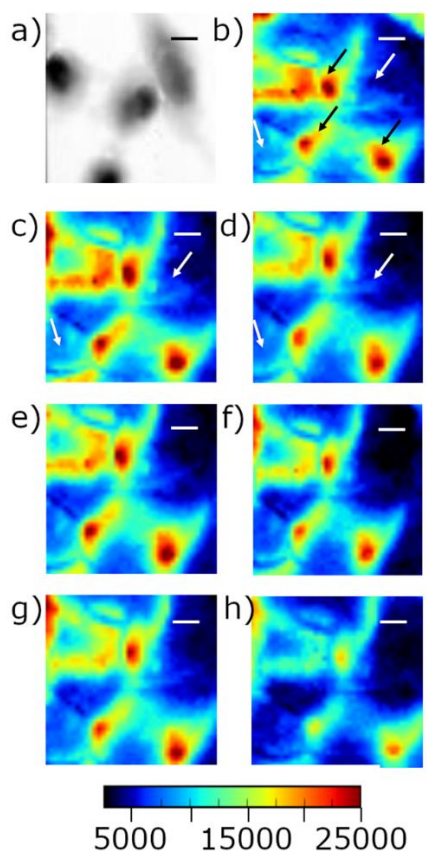

**Figure 4**: a) Absorption image of A549 cells deposited on a SiN membrane; XRF evolution map of Oxygen content in the analyzed GrELC, recorded at time b) 0 hours, c) 2.7 hours, d) 5.2 hours, e) 7.7 hours, f) 10.4 hours, g) 17 hours, h) after chamber venting and pumping. The maps were acquired at 1.03 keV, in scanning mode with a step size of 800 nm and 5 s/pixel acquisition time for XRF spectra, over an area of 32 µm × 32 µm. Scale bar is 5 µm. The color scale represents the count rate.

Overall, the presented FTIRM and LE-XRF results reveal that GrELC are able to contain very thin water layers for many hours, even in-vacuum environments, and that the water layer stands above the cells, due to the conformability of the graphene layers. The water retention time both at atmospheric ambient and HV conditions is greater than 14 hours, accordingly to FTIRM and LE STXM-XRF data. In addition, thanks to the very thin layer of water confined onto the cells in our GrELCs, our system combines the advantages to work in physiological conditions, while providing a spectral quality comparable to that of dried samples.

So far, the measurements were dealing with pristine graphene encapsulated objects. We now explore the opportunities provided with an additional hydration source such as hydrogel containing encapsulants. To quantitatively monitor the ability of GrELC to retain the water inside hydrated encapsulated objects we used PEGDA/PBS solution mixed with (ca 0.1%) of fluorescein. Fluorescein is a water-soluble fluorophore that has a broad fluorescence band around 532 nm in aqueous solutions when exited with UV light, but its fluorescence is quenched in a dry state and, therefore, can be used as optical hydration marker. Fluorescence microscopy was used to monitor the fraction of water inside of hydrogel media as a function of time when GrELC with graphene encapsulated hydrogel micro-sample was exposed to low vacuum (ca $2\cdot10^3$ Pa). For that the GrELCs were placed inside a specially designed vacuum chamber with an optical viewport and observed by the fluorescence microscope during pumping. The fluorescence intensity curves are displayed within Figure 5 for the time interval t=0 to 6 hours. The intensity of the fluorescence signal from every hydrogel sample was laterally averaged and normalized to its original value at t=0. Figure 5 shows the fluorescence intensity curves averaged over six GrELCs that has 'uncoated' control hydrogel samples (blue curve) exposed to vacuum and those that were encapsulated by graphene membranes (red curve) under the same conditions. The control samples had water evaporating at an expedient rate when pumped inside the vacuum chamber, whereas the graphene covered cells had their water content preserved for many hours, exhibiting a retention time greater than 6 hours, while it is about 3 hours for the control. The retention time increases to nearly 200 hours for encapsulated hydrogel exposed to air (50% HR, 26 C), in comparison to about 48 hours for controls in the same experimental conditions (See Figure S4). Furthermore, the comparison of water retention times for the two sample sets further confirm that the graphene membrane effectively preserves the hydration of the encapsulated sample in low vacuum. Several factors can be responsible for dehydration rates of the encapsulated objects. These are interfacial diffusion of water molecules between graphene and $SiO_2$/SiN layer (including the wrinkle structures) and through the defects in the graphene layer itself. The interfacial transports from the hydrated sample occurs laterally outward, towards the edges of the lid, where it readily evaporates into the vacuum. Furthermore, any local or extended defects along the graphene membrane, such as tears or wrinkles, would facilitate the diffusion and contribute to evaporation rates. Our results demonstrate that the conformal encapsulation of the graphene around the samples was both unbroken and isolating from the outside environment and that the drying of the samples would be primarily associated with interfacial diffusion. Summarizing, graphene was able to preserve water content inside the model gel structures for timescales suitable to study biological samples with a wide array of analytical techniques and yield viable signal to noise ratios.

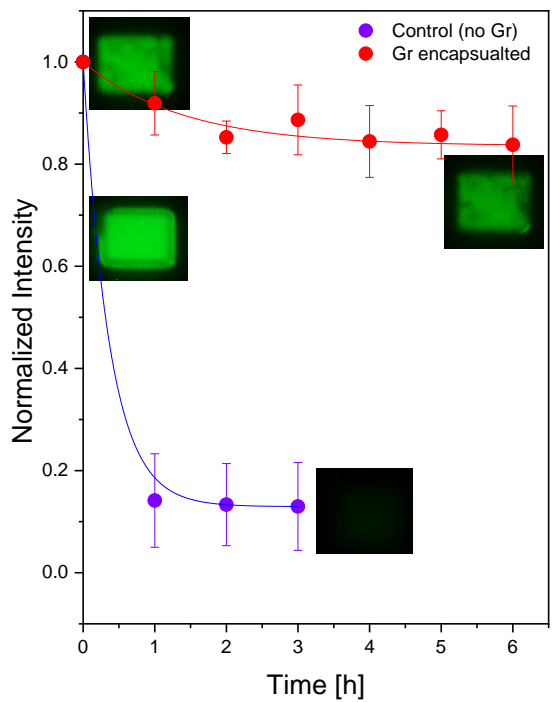

**Figure 5**: Normalized fluorescence intensity of die impregnated hydrogel microsamples as a function of time for uncovered PEGDA gelated samples (blue curve) and graphene encapsulated gel samples in low vacuum (red curve). Every point is an average of six samples and has a corresponding error bar. Insets: the corresponding fluorescence microscopy images in the beginning and in the end of the cycle. The width of the fields of view is 20 microns

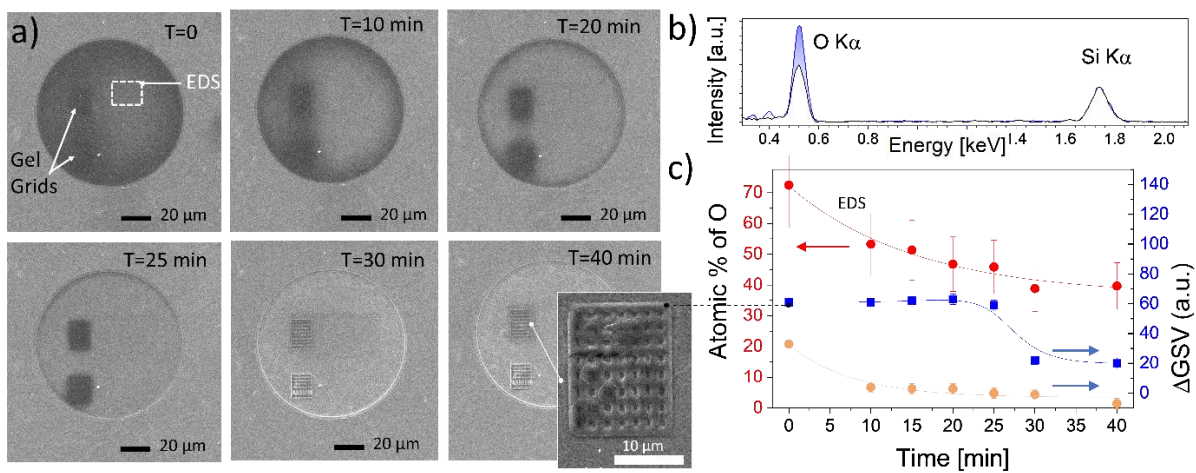

**Figure 6**. a) SEM images of water encapsulated by graphene overtop of patterned gel lattice as the water evaporated Inset: graphene encapsulated hydrated gel lattice structure. b) EDS spectra recorded from the dashed square in (a) at t=0 min and t=40 min. c) Red curve: the evolution of atomic percentage of oxygen with time determined from EDS. The data are fit as single exponent decay with a time constant τ ≈ 16 min. The differences between the SEM gray scale values of the $SiO_2$ frame and gel lattice structure (blue curve) and EDS region in (a) (orange curve) as a function of time

To assess the effect that hydrogels have on internal water retention, SEM and EDS were preformed of a graphene encapsulated patterned gel structure that mimics a hydrated biological object under high vacuum (ca. $10^{-4}$ Pa). Figure 6(a) displays SEM images of the graphene encapsulated hydrated SiN viewport which contained two gel lattice structures patterned by a 5 keV electron beam. The sample has been under vacuum for 1 hour prior to the collection of the first image in the Figure 6(a) and still bulk water can clearly be observed as the dark areas due to (i) lower electron yield and (ii) attenuated SE signal from SiN membrane. SEM images were collected periodically until the sample's appearance stabilized over a period of 40 minutes. The EDS provided elemental composition of the selected (dashed square in the Figure 6a) during SEM imaging and was used to quantify the water content entrapped by the graphene membrane as a function of time. The EDS spectra displayed the characteristic Kα peaks of O, and Si, at 525 eV and 1740 eV emission energies respectively (Figure 6b). The atomic concentration of O from the water region was calculated as a function of time (Figure 6c). As can be seen, the oxygen content is greater at the onset of the experiment and this is consistent with the presence of molecular water slowly leaking into the vacuum with time. The similar results can be obtained from the comparative analysis of the temporal evolution of SEM gray scale values (GSV) of the graphene encapsulated SiN and gel lattice structure (blue and orange curves in the Figure 6c). The graphene encapsulated porous hydrogel structures were able to retain water significantly longer (at least 20 min) compared to the surrounding SiN window as their GSV remained darker as the surrounding SiN window dried out. Whether the evaporation is complete, or a residual water content persists in the test structures is hard to determine. Nevertheless, the SEM experiments here presented indicate that the hydrophilic and porous properties of the crosslinked gel would be beneficial for extending the liquid retention time in the hydrated samples also in high vacuum conditions.

**Conclusions**

The fabrication and tests of a multi-technique accessible platform for biological material encapsulation via a thin graphene membrane was described using test samples made of hydrogel microstructures, yeast and A549 cells. The proposed GrELC does not use the standard PMMA-based graphene encapsulation procedure which can potentially affect biological specimens. Instead, it employs the microfabricated thin flexible polymeric lid with graphene covered viewports for conformal encapsulation. We show that GrELCs have a wide range of applications, being suitable for both photon (IR and soft X-rays) and electron based spectromicroscopies, and offer multiple identical viewports that can be used for combinatorial studies. The platform was sought to retain water for period of times sufficient for the selected analytical techniques and corresponding environments, that spans from ambient pressure conditions (FTIRM) to HV (SEM-EDS and LE-XRF). The hydration retention times measurement indicate that the GrELCs provide physiologically relevant water around the cells even in HV environment for a time sufficient for LE-XRF experiment (typically a few hours for a single cell map). A strategy for enhancing water retention has been tested by adding an additional PEGDA hydrogel nanoporous material as a co-encapsulant. Towards this goal, a patternable gel immobilization technology was developed to interface biological objects with high fidelity using with electron beam induced radiolysis as a local crosslinking agent. This method can be further developed to incorporate engineered fluidic channels for feeding and supporting live cells also when a direct hydrogel immobilization is not allowed as well as for patterning analytical probes such as electrodes to study local electrophysiology in operando mode. The idea to decouple graphene scooping procedure and sample encapsulation procedures is indeed a key point for dissemination of the graphene encapsulation method to a variety of other spectroscopic and microscopic techniques for *in-situ* characterization of diverse phenomena.

**Author Contributions**

A.K. and L.V. conceived the idea and directed the work. Chip general design was developed by A.K. L.V. A.M. and H.G. Chip fabrication and process development was done by H.G. C.A. conducted SEM and fluorescence tests and related data analysis. A.M. A.G. and G.B. carried out synchrotron radiation-based chip tests and spectroscopic measurements. The manuscript was compiled by C.A. and A.K. using contributions of all authors. All authors have given approval to the final version of the manuscript.

**Conflicts of interest**

There are no conflicts to declare.

**Disclaimer**

Any mention of commercial products in this manuscript is for information only; it does not imply recommendation or endorsement by NIST

**Acknowledgements**